\documentclass{article}
\usepackage{spconf,amsmath,epsfig}

\usepackage{caption}
\usepackage{subcaption}
\usepackage{graphicx}
\usepackage{enumitem}
\usepackage{xcolor}
\usepackage{amsmath,amssymb}
\usepackage{hyperref}
\usepackage{multirow}
\usepackage{hhline}
\usepackage{tabularx}
\usepackage{enumitem}
\usepackage{appendix}
\usepackage{float}
\usepackage{multirow}
\usepackage{hhline}
\usepackage{tabularx}

\hypersetup{
    colorlinks=true,
    linkcolor=blue,
    filecolor=blue,      
    urlcolor=blue,
    pdftitle={Overleaf Example},
    pdfpagemode=FullScreen,
    }
\title{Structured Pruning and Quantization for Learned Image Compression}

\name{Md Adnan Faisal Hossain \qquad Fengqing Zhu}
\address{Elmore Family School of Electrical and Computer Engineering,\\Purdue University, West Lafayette, Indiana, U.S.A.}

\begin{document}\sloppy
%
\maketitle
\begin{abstract}
The high computational costs associated with large deep learning models significantly hinder their practical deployment. Model pruning has been widely explored in deep learning literature to reduce their computational burden, but its application has been largely limited to computer vision tasks such as image classification and object detection. In this work, we propose a structured pruning method targeted for Learned Image Compression (LIC) models that aims to reduce the computational costs associated with image compression while maintaining the rate-distortion performance. We employ a Neural Architecture Search (NAS) method based on the rate-distortion loss for computing the pruning ratio for each layer of the network. We compare our pruned model with the uncompressed LIC Model with same network architecture and show that it can achieve model size reduction without any BD-Rate performance drop. We further show that our pruning method can be integrated with model quantization to achieve further model compression while maintaining similar BD-Rate performance. We have made the source code available at \href{https://gitlab.com/viper-purdue/lic-pruning.git}{gitlab.com/viper-purdue/lic-pruning}.

\end{abstract}
\begin{keywords}
learned image compression, model compression, pruning, neural architecture search, quantization
\end{keywords}

\section{Introduction}
\label{sec:intro}
\vspace{-0.2cm}

\begin{figure}[t]
 \centerline{\epsfig{figure=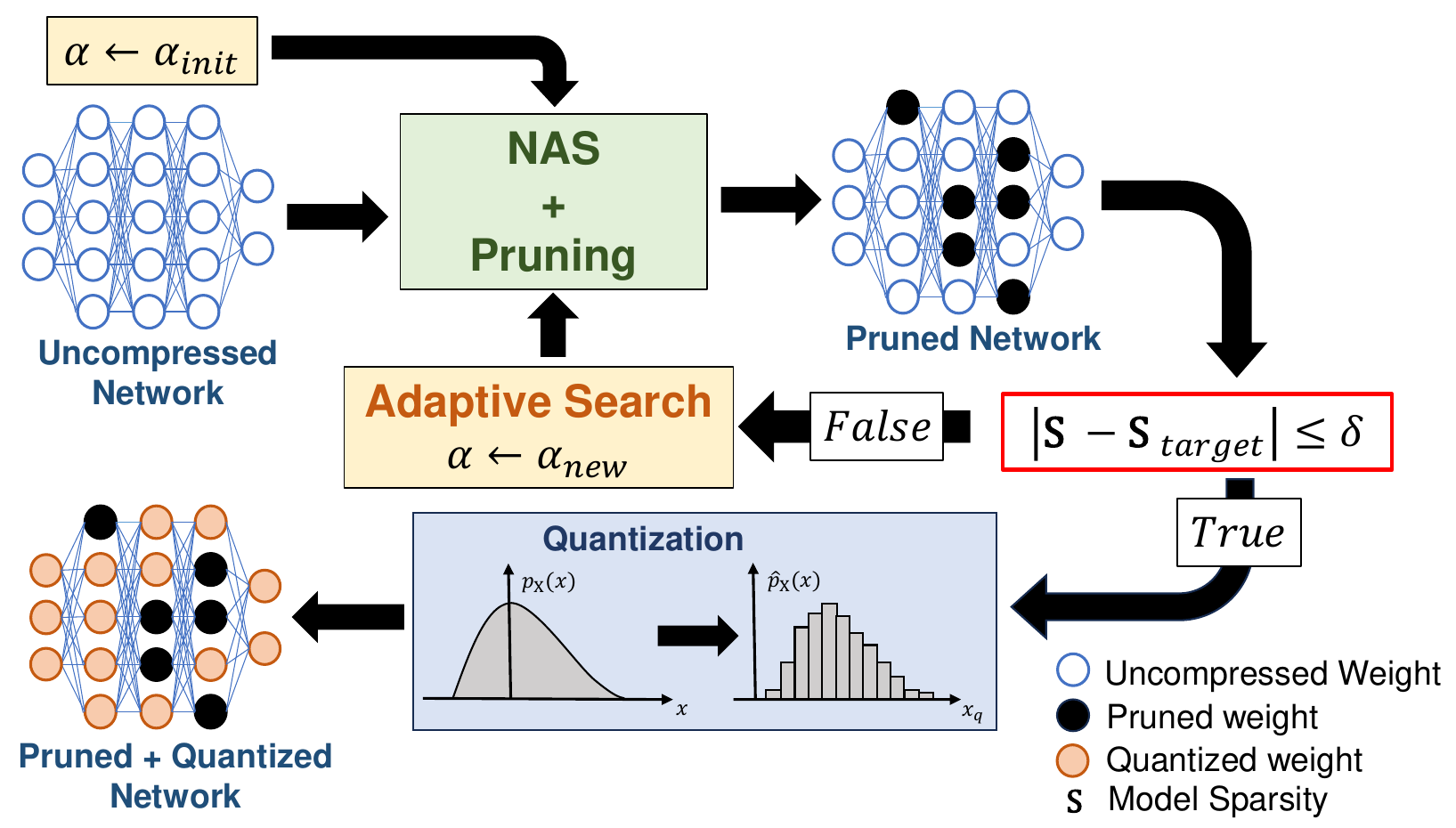, width=1.0\linewidth}}
\vspace{-0.2cm}
\caption{Overview of our proposed structured pruning and quantization method. Starting with an initial value of the parameter $\alpha$, we obtain the pruned LIC model from the uncompressed model. If the sparsity $S$ of the pruned model is within a threshold $\delta$ of the desired sparsity $S_{\textit{target}}$, quantization is performed. Otherwise, an adaptive search algorithm adjusts $\alpha$, and the pruning process is repeated.}
\label{fig:overview}
\vspace{-0.5cm}
\end{figure}

Learned Image Compression (LIC) \cite{balle2018variational, minnen2018joint, duan2023lossy, duan2023qarv, jiang2023mlic} has recently shown immense progress surpassing the bounds of traditional codecs such as JPEG \cite{wallace1991jpeg} and VTM \cite{bross2021overview}. However, LICs utilize deep learning models which have large memory requirements and latency constraints, hindering their deployment on resource-limited hardware. To address these challenges, we propose to utilize structured pruning and quantization for the compression of LIC models. While there have been existing works that investigate the quantization of LIC models \cite{hong2020efficient, sun2020end, sun2021learned, shi2023rate, jeon2023integer}, the use of pruning for compression of LIC models is limited to only the decoder or hyperprior model \cite{luo2022memory, liao2022efficient, munna2023complexity}. Furthermore, due to the complementary nature of quantization and pruning, it can be beneficial to perform them both simultaneously as discussed in \cite{bai2023unified}. Therefore, in this work, we develop a joint pruning and quantization method for the compression of LIC models.


An important aspect of structured pruning \cite{he2019filter, lin2020hrank, tang2020scop, sui2021chip} involves the choice of pruning granularity; filter pruning or filter channel pruning. Specifically, filter pruning and filter channel pruning correspond to pruning along the output dimension and input dimension of a network layer, respectively. While previous works \cite{lebedev2016fast, meng2020pruning, ma2020pconv} focus on the effectiveness of different pruning granularities in isolation, we show that simultaneously pruning filters and filter channels can further enhance performance. 

Due to the large search space involved, manually tuning the layer-wise pruning ratio is cumbersome in practice. Instead, we propose a Neural Architecture Search (NAS) method to search for the layer-wise pruning ratios. Our method uses the rate-distortion loss for image compression as the search criterion and aims to find the best distribution of pruning ratios while conforming to model size constraints. After pruning the LIC, we quantize our pruned model to fixed precision weights and perform joint finetuning to achieve enhanced model compression. An overview of our proposed method is shown in Fig.~\ref{fig:overview}. Our main contributions are summarized as follows,
\setlist{nolistsep}
\begin{itemize}[noitemsep]
  \item We propose a structured pruning method for Learned Image Compression (LIC) models based on pruning both filters and filter channels.
  \item We propose a two-stage Neural Architecture Search (NAS) for determining the layer-wise pruning ratios under the constraints of a given model size.
  \item We propose simultaneous structured pruning and quantization of LIC models to achieve enhanced model compression. 
\end{itemize}

\vspace{-0.3cm}
\section{Preliminaries}
\label{sec:prelim}
\vspace{-0.2cm}

\subsection{Learned Image Compression}
\label{sec:prelim_1}
\vspace{-0.2cm}

Learned Image Compression is a form of image source coding that operates in the transform coding paradigm. This involves transformation of the input image into a latent representation which is subsequently quantized and entropy-coded into bit-streams for transmission or storage using a prior distribution. In our work, we adopt a variant of the Mean Scale Hyperprior LIC model from \cite{minnen2018joint} to perform the compression task. The compression network consists of a core encoder $g_a$ and decoder $g_s$ as well as a hyper encoder $h_a$ and hyper decoder $h_s$. The input image $x$ is passed through $g_a$ to generate the latent representation $y=g_a(x)$. $y$ is then passed through $h_a$ to produce the hyper latent $z=h_a(y)$ which is subsequently quantized to $\hat{z}$ and entropy coded using a factorized entropy model $p_{\hat{z}}(\hat{z})$. $\hat{z}$ is recovered at the hyper decoder $h_s$ and used to generate $\mu$ and $\sigma$ which are the parameters of the parametric prior distribution $p_{\hat{y}}(\hat{y}; \mu,\sigma)$ that is used to perform entropy coding of the quantized latent representation $\hat{y}$. The quantized, compressed representation $\hat{y}$ is losslessly decoded at the decoder $g_s$ and used to reconstruct the image $\hat{x}=g_s(\hat{y})$. 
The network is trained using the Rate-Distortion loss ($L_\text{RD}$) shown in Eq. \eqref{eqn:1} where the Lagrangian multiplier $\lambda$ is varied to trade-off between the rate and distortion term:

\vspace{-0.5cm}
\begin{equation}  \label{eqn:1}
\begin{split}
L_{\text{RD}} & = \text{Rate} + \lambda \cdot \text{Distortion} \\
& = \mathbb{E}_{X \sim p_x} [-\log_{2}p_{\hat{y}|z}(\hat{y}|z) - \log_{2}p_{\hat{z}}(\hat{z})] + \lambda \cdot d(x, \hat{x})  .
\end{split}
\end{equation}
\vspace{-0.9cm}

\begin{figure*}[t]
\centerline{\epsfig{figure=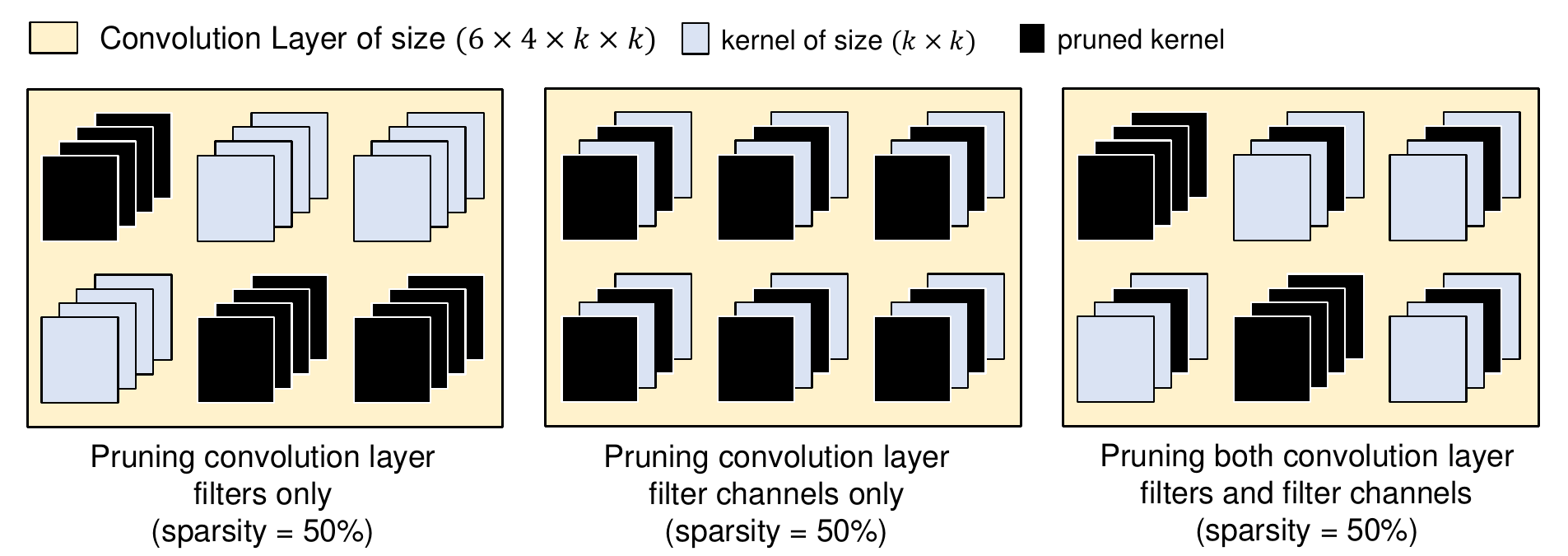, width=0.85\textwidth}}
\vspace{-0.2cm}
\caption{Convolution layer pruned to the same sparsity using different granularity. The figure on the left shows filter pruning, the middle one shows channel pruning and the right one displays our proposed pruning of filters + filter channels.}
\vspace{-0.2cm}
\label{fig:pruning_granularity}
\vspace{-0.2cm}
\end{figure*}

\subsection{Neural Network Pruning}
\label{sec:prelim_2}
\vspace{-0.2cm}

Pruning aims to remove redundant weights from neural networks based on a predefined pruning criterion and pruning ratio. It can be primarily divided into two categories based on the pruning granularity: unstructured pruning \cite{han2015learning} and structured pruning \cite{molchanov2019importance}. Unstructured pruning (or weight pruning) focuses on selecting the weights to be pruned within a layer based on the individual importance of each weight element. On the contrary, structured pruning removes entire filters from a layer based on an average importance criterion measured using all the weight elements of a particular filter. In our work, we perform structured pruning since weight pruning leads to unstructured sparsity patterns which require the use of complicated hardware support to achieve model compression. The pruning criterion refers to the metric applied in determining the relative importance of filters within a neural network layer. Primarily, the pruning criterion is either filter-dependent such as filter norm \cite{he2018soft} or feature based like average rank of activation feature maps \cite{lin2020hrank}. Recent work on feature guided filter pruning \cite{lin2020hrank, tang2020scop, sui2021chip} has shown that they are capable of achieving better performance compared to filter guided counterparts as the information content of the feature maps provide better guidelines for determining the importance of filters than the characteristics of the filter weights themselves. 

\vspace{-0.2cm}
\subsection{Neural Architecture Search}
\label{sec:prelim_3}
\vspace{-0.2cm}

Neural Architecture Search (NAS) involves searching for the optimal neural network architecture with reinforcement learning \cite{he2018amc}, evolutionary algorithms \cite{liu2019metapruning} or gradient-based methods \cite{guo2020dmcp}. This involves modeling the pruning process as a method for identifying the best sub-network structure rather than finding the most important weights of the network \cite{lin2019towards}. In short, NAS aims to replace manual hyper-parameter or network architecture tuning of neural networks with an automated process based on an overall optimization criterion. In our work, we devise a NAS method for determining the layer-wise pruning ratio of our pruned LIC model. 

\vspace{-0.1cm}
\subsection{Model Quantization}
\label{sec:prelim_4}
\vspace{-0.2cm}

Model quantization primarily consists of Post-Training Quantization (PTQ) \cite{nagel2020up, wei2022qdrop, shi2023rate} and Quantization-Aware Training (QAT) \cite{choi2018pact, esser2019learned, lee2021network}. Although PTQ only requires a small number of unlabeled calibration data and no retraining making it suitable for off-the-shelf deployment, QAT offers better model performance. Majority of the work on QAT simultaneously trains the quantized neural network over a task loss alongside the quantization error loss. Most approaches for training low precision networks adopt uniform quantizers which involves scaling the dynamic range of the weights and then quantizing to integers. The scaling is performed using quantization scale and bias parameters, respectively, and the bounds of the new dynamic range is determined by the quantization bit-width $b$. In general, the quantization bit-width is inversely related to the achieved model compression, and the bit-width can be varied to obtain a trade-off between model compression and rate-distortion performance.

\vspace{-0.3cm}
\section{Method}
\label{sec:method}
\vspace{-0.2cm}

In this section, we first introduce our method for simultaneously pruning filters and filter channels. Next, we demonstrate our proposed NAS procedure for determining the layer-wise pruning ratio of the pruned model. Finally, we outline the joint pruning and quantization scheme. 

\vspace{-0.2cm}
\subsection{Pruning Filters + Filter Channels}
\label{sec:method_1}
\vspace{-0.2cm}

Due to the superiority of feature-guided pruning over filter-guided pruning, as discussed in Sec.~\ref{sec:prelim_2}, we choose to model our pruning formulation as a feature-guided pruning method. We define structured pruning primarily as either pruning filters, which corresponds to pruning using the \textbf{output feature maps} of a neural network layer, or pruning filter channels, which corresponds   to pruning using the \textbf{input feature maps}.

Consider the $l$-th convolutional layer of a neural network having a set of filter weights $\text{\boldmath$W$}^l=\{F_1^l, F_2^l, ... F_{c_l}^l\} \in R^{c_l \times c_{l-1} \times h \times w}$. The input feature maps are $\text{\boldmath$A$}^{l-1}=\{A_1^{l-1}, A_2^{l-1}, ... A_{c_{l-1}}^{l-1}\} \in \mathbb{R}^{c_{l-1} \times h \times w}$ and the output feature maps are $\text{\boldmath$A$}^{l}=\{A_1^{l}, A_2^{l}, ... A_{c_{l}}^{l}\} \in \mathbb{R}^{c_{l} \times h \times w}$. The $i$-th filter of $\text{\boldmath$W$}^l$, $F_i^l$ is convolved with $\text{\boldmath$A$}^{l-1}$ to generate the $i$-th channel of the output feature map $A_i^l$. The task of output feature map guided filter pruning involves finding the set of least important feature maps $\text{\boldmath$A$}^{(l, \kappa)} \in \text{\boldmath$A$}^l$ based on a predefined channel importance criterion and pruning ratio $\kappa$. The pruning ratio determines the number of elements in $\text{\boldmath$A$}^{(l, \kappa)}$ using the relationship $\|\text{\boldmath$A$}^{(l, \kappa)}\| = \kappa \cdot \|\text{\boldmath$A$}^l\|$ where $\| \cdot \|$ defines the number of elements in a set. Using the mapping between the filters of $\text{\boldmath$W$}^l$ and the channels of $\text{\boldmath$A$}^l$, the final step involves determining the set of pruned filters $\text{\boldmath$W$}^{(l,\kappa)} \in \text{\boldmath$W$}^l$ corresponding to the elements of the set of least important feature maps $\text{\boldmath$A$}^{(l,\kappa)}$. 

Identifying the pruned filters, $\text{\boldmath$W$}^{(l, \kappa)}$, using only the output feature maps is sub-optimal as it only exploits the redundancy between filters of a particular layer, and ignores the redundancy within the channels of each filter $F_{i}^l$. Therefore, we propose to perform a sequential pruning of the filter weights by first considering the output feature maps $\text{\boldmath$A$}^l$ as described previously, and then the input feature maps $\text{\boldmath$A$}^{l-1}$. We design the following process to prune the filter channels using input feature maps. The $k$-th input feature map $A_k^{l-1}$ is element-wise multiplied with the $k$-th channel of each of the filters $F_i^l$ in $\text{\boldmath$W$}^l$ during convolution, where $i \in \{1,2 ... c_l\}$. We define a set $\text{\boldmath$\theta$}^l=\{\theta_1^l,\theta_2^l ... \theta_k^l ... \theta_{c_{l-1}}^l \} \in \mathbb{R}^{c_{l-1} \times c_l \times h \times w}$ where each element $\theta_k^l$ contains the $k$-th channel of all the filters in $\text{\boldmath$W$}^l$. Similar to pruning based on output feature maps, we first determine a subset of the input feature maps $\text{\boldmath$A$}^{(l-1,\kappa)}$ under the same pruning criterion and pruning ratio. Using this subset and the mapping between input feature maps and filter channels, we determine a subset of $\text{\boldmath$\theta$}^l$ that corresponds to the pruned channels in each of the filters of $\text{\boldmath$W$}^l$. Therefore, by sequentially pruning the $l$-th convolution layer $\text{\boldmath$W$}^l$, based on output feature maps and then input feature maps, we can first remove redundant filters from $\text{\boldmath$W$}^l$ and then remove redundant channels from each of the remaining filters of $\text{\boldmath$W$}^l$. Our proposed method for simultaneously pruning filters and filter channels is summarized in Fig.~\ref{fig:pruning_granularity}.  

\vspace{-0.2cm}
\subsection{NAS for Layer-wise Pruning Ratio}
\label{sec:method_2}
\vspace{-0.2cm}

\begin{figure*}[t]
\centerline{\epsfig{figure=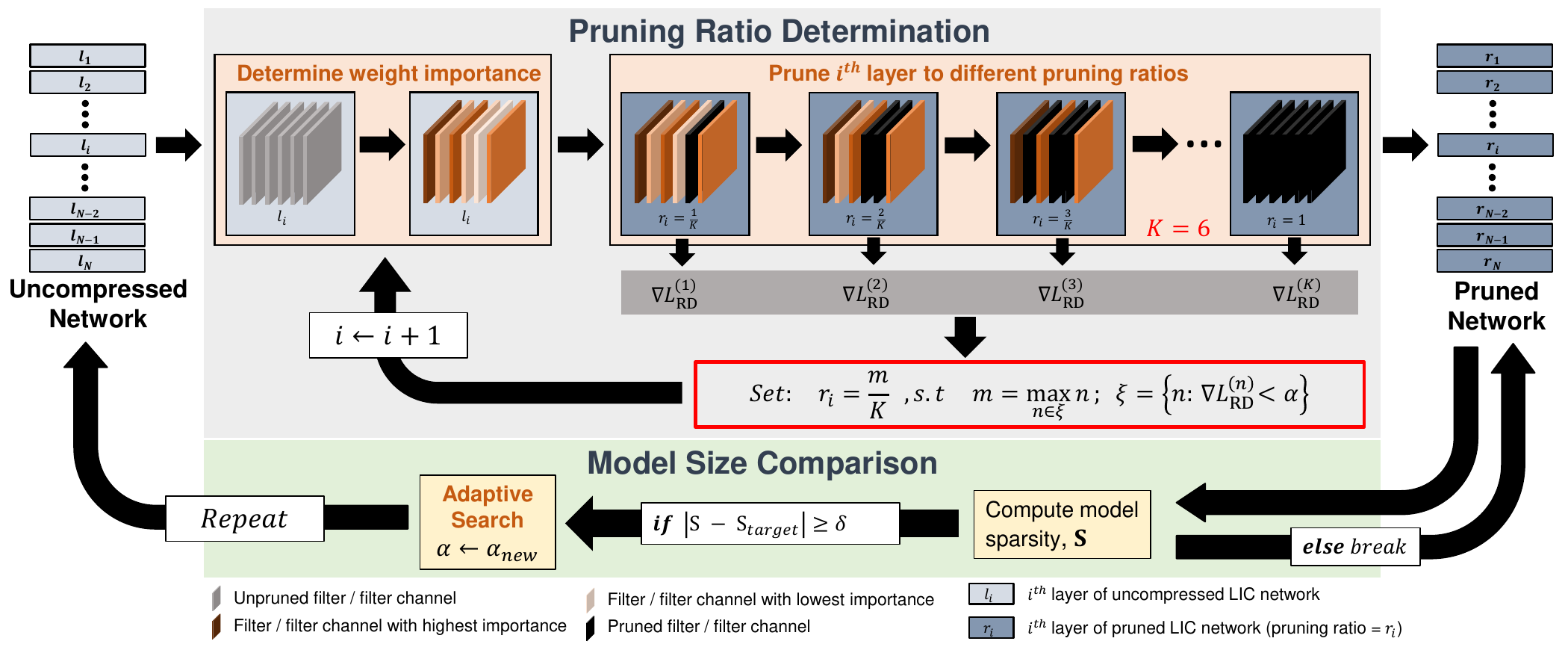, width=0.95\textwidth}}
\vspace{-0.1cm}
\caption{Proposed two-stage Neural Architecture Search (NAS) method for determining the layer-wise pruning ratio conforming to a fixed model sparsity constraint. In the first stage (highlighted in gray), the layer-wise pruning ratio is determined, and in the next stage (highlighted in green), the model size comparison is performed. If the desired sparsity is achieved, the process terminates, otherwise the search goes to the next iteration with a new $\alpha$. $\nabla{L_{\text{RD}}}$, $\alpha$, $\delta$, $K$, $S$ and $S_{\text{target}}$ are defined in Sec.~\ref{sec:method_2}.}
\vspace{-0.2cm}
\label{fig:NAS}
\vspace{-0.2cm}
\end{figure*}

We propose a two-stage NAS method to determine the layer-wise pruning ratio as illustrated in Fig.~\ref{fig:NAS}. Our method operates on each layer of the neural network sequentially, going from top to bottom. The first stage of our process consists of determining the pruning ratio for each layer based on the change in rate-distortion loss $\nabla{L_{\text{RD}}}$ and a threshold of tolerance $\alpha$. In the second stage, we measure the sparsity $S$ of the pruned model and match it against a target sparsity $S_{\text{target}}$. In the case where $S$ is close $S_{\text{target}}$, we terminate the process obtaining our layer-wise pruning ratios. Otherwise, we increment $\alpha$ and repeat the entire process.

For each layer, the first step of our iterative process consists of determining the order of importance of filters based on a predetermined pruning criterion. The filters are then progressively pruned (pruning ratio is sequentially increased) starting from the least important one, and on each pruning step $\nabla{L_{\text{RD}}}$ is measured and stored. We then select the highest pruning ratio that achieves $\nabla{L_{\text{RD}}} \leq \alpha$ as the pruning ratio for that layer. The whole network is also finetuned on a small calibration dataset on each step before calculating $\nabla{L_{\text{RD}}}$, as the network performance drops significantly after pruning. While determining the pruning ratio of a particular layer, all other layers are kept uncompressed with their original weights. Mathematically, we can define the pruning ratio $r_i$ of the $i$-th layer with $M$ filters obtained from our search as:
\begin{equation}  \label{eqn:2}
r_i = \frac{m}{M}, \text{s.t.}  \hspace{0.4cm} m = \max_{n \in \xi} n ;\hspace{0.2cm} \xi = \{n:\nabla{L_{RD}^{(n)}} < \alpha \}
\end{equation}
where $m$ is the number of pruned filters and $\nabla{L_{RD}^{(m)}}$ is the change in rate-distortion loss due to pruning $m$ out of the $M$ filters of the $i$-th layer.

To facilitate faster computation instead of pruning one filter in each step, we prune the filters in groups of size $K$. A larger value of $K$ lowers the computation time while a smaller value can achieve more accurate pruning choices. We use an identical method to also prune the filter channels by first measuring the importance of each filter channel, then progressively pruning and finetuning them before measuring the $\nabla{L_{\text{RD}}}$ and assigning a pruning ratio.

The first stage of our NAS process can determine the layer-wise pruning ratios for a given value of $\alpha$. In the second stage, we manipulate the value of $\alpha$ so that the sparsity $S$ of our pruned model is equal to our desired target sparsity $S_{\text{target}}$. Therefore on each iteration of our search, after the layer-wise pruning ratios are determined we compute $S$, and if $|S - S_{\text{target}}| \leq \delta$, we terminate the search and conclude that the target sparsity has been achieved. Otherwise, we increment alpha using an adaptive search algorithm and perform another iteration of our two-stage search. The adaptive search algorithm is modeled as a variable step size search that adaptively changes the increment applied to $\alpha$ based on the value of $|S - S_{\text{target}}|$. Therefore, if the difference between $S$ and $S_{\text{target}}$ is small, smaller increments are applied, and if the difference is large, larger increments are applied.

\vspace{-0.2cm}
\subsection{Joint Pruning and Quantization}
\label{sec:method_3}
\vspace{-0.2cm}

Pruning LIC models to high sparsity values can lead to significant reduction in rate-distortion performance as the network might no longer have sufficient parameters to generate strong representations of the input. To tackle this, we suggest pruning the model to a moderate sparsity level and then achieving additional model compression through weight quantization. We show that by converting the pruned model parameters to a lower precision (\textit{float32} to \textit{int8}) we can achieve better rate-distortion performance under the same model size constraints than by only pruning the model. We describe our model compression scheme as a joint pruning and quantization method designed particularly for LIC models.

We adopt uniform quantization where we scale the dynamic range of the weights and then quantize them to integers using $b$ bits. We quantize both the network weights, $w$ and activations, $a$ using the pair of quantization parameters $(s_w, z_w)$ and $(s_a, z_a)$, respectively. The quantization operations for $w$ and $a$ are defined in Eq.~\eqref{eqn:3} and Eq.~\eqref{eqn:4}, respectively.  
\begin{equation}  \label{eqn:3}
\hat{w} = s_w \times \left\{\left\lfloor \text{clip}\left( \frac{w}{s_w}+z_w; 0, 2^{b}\right)  \right\rceil - z_w \right\} 
\end{equation}
\begin{equation}  \label{eqn:4}
\hat{a} = s_a \times \left\{\left\lfloor \text{clip}\left( \frac{a}{s_a}+z_a; -2^{b-1}, 2^{b-1}-1\right)  \right\rceil - z_a \right\} 
\end{equation}
Here $\lfloor \cdot \rceil$ refers to the integer rounding operation and the scaled weights are clipped between $[L, U]$. We quantize the weights to unsigned integers and the activations to signed integers as can be seen from Eq.~\eqref{eqn:3} and Eq.~\eqref{eqn:4}. We also utilize a channel-wise quantization scheme whereby each filter of a convolution layer has its own distinct set of quantization parameters.

\begin{table}[t]
\begin{center}
\caption{BD-Rate performance of our proposed pruning method across different pruning criterion using the Mean Scale Hyperprior \cite{minnen2018joint} LIC model.} \label{tab:tab1}
\vspace{-0.3cm}
\scalebox{0.90}{
\begin{tabular}{c|c|c|c}
\hline
Pruning Type & \begin{tabular}[c]{@{}c@{}}Pruning\\ Criterion\end{tabular} & \begin{tabular}[c]{@{}c@{}}BD-Rate\\  Kodak \\ (\%)\end{tabular} & \begin{tabular}[c]{@{}c@{}}Parameter \\ Reduction \\ (\%)\end{tabular} \\
\hline
None & None  & 0 & 0 \\
\hline
Filters & \multirow{3}{*}{\begin{tabular}[c]{@{}c@{}}\\ L2-Norm\\ \cite{li2016pruning} \end{tabular}}   & +42.19 & 30.5 \\
\hhline{-~--}
Filters + Filter Channels &  & +30.56 & 30.2 \\
\hhline{-~--}
\begin{tabular}[c]{@{}c@{}}Filters + Filter Channels\\ + NAS (Proposed)\end{tabular} &  & +10.33 & 30.2 \\
\hline
Filters & \multirow{3}{*}{\begin{tabular}[c]{@{}c@{}}\\ HRANK\\ \cite{lin2020hrank} \end{tabular}}   & +34.09 & 30.3 \\
\hhline{-~--}
Filters + Filter Channels &  & +7.83 & 29.9 \\
\hhline{-~--}
\begin{tabular}[c]{@{}c@{}}Filters + Filter Channels\\ + NAS (Proposed)\end{tabular} &  & +0.37 & 30.1 \\
\hline
Filters & \multirow{3}{*}{\begin{tabular}[c]{@{}c@{}}\\ CHIP\\ \cite{sui2021chip}\end{tabular}}   & +7.17 & 30.3 \\
\hhline{-~--}
Filters + Filter Channels &  & -1.73 & 29.9 \\
\hhline{-~--}
\begin{tabular}[c]{@{}c@{}}Filters + Filter Channels\\ + NAS (Proposed)\end{tabular} &  & -2.00 & 29.8 \\ \hline
\end{tabular}}
\vspace{-0.7cm}
\end{center}
\end{table}

Once we determine the layer-wise pruning ratios using the method from Sec.~\ref{sec:method_2}, we sequentially prune and quantize the model. We then perform finetuning of the pruned + quantized model. To represent the pruned weights during training, we utilize static pruning masks. During training, we perform gradient updates on the full-precision weights and use the quantized weights to do the forward propagation. We set the quantization parameters for weights $s_w$ and $z_w$ as learnable parameters that are trained but remain static during inference. However, the quantization parameters for activations $s_a$ and $z_a$ are determined dynamically during inference as shown in \cite{shi2023rate}. We choose the rate-distortion loss from Eq.~\eqref{eqn:1} as the optimization function for this entire process.

\vspace{-0.3cm}
\section{Experiments}
\label{sec:experiments}
\vspace{-0.2cm}

We compare the performance of our proposed pruning methods using three different pruning criteria, L2-Norm \cite{li2016pruning}, HRANK \cite{lin2020hrank} and CHIP \cite{sui2021chip}. Next, we show the coding performance of our proposed pruning methods across different model sizes. Finally, we discuss the performance gain achieved by performing quantization in conjunction with pruning. 

\vspace{-0.2cm}
\subsection{Experimental Setup}
\label{sec:exp_1}
\vspace{-0.2cm}

We obtain the uncompressed baseline LIC model by training a variant of the Mean-Scale Hyperprior model from \cite{minnen2018joint}. We train the model using the COCO dataset \cite{lin2014microsoft} for $90$ epochs using the Adam optimizer and a batch size of $16$. We also utilize a cosine learning rate decay with an initial learning rate of $10^{-4}$. All images are randomly cropped to patches of $256 \times 256$ during training. We train five different baseline networks to obtain five different quality levels of compression corresponding to $\lambda=\{0.0018, 0.0035, 0.0067, 0.0130, 0.0250\}$. 

We measure the rate-distortion performance of our pruned models and quantized models using the Bjontegaard delta rate (BD-Rate) \cite{bjontegaard2001calculation} with respect to the baseline LIC model. We measure the model size reduction of the pruned models by comparing the percentage parameter reduction with respect to the baseline, and we measure the model size reduction achieved by the pruned+quantized models by calculating their model size in MB. We use the Kodak \cite{kodak1993kodak} dataset to evaluate the BD-Rates.

\vspace{-0.2cm}
\subsection{Coding Performance of Proposed Pruning Methods}
\label{sec:exp_2}
\vspace{-0.2cm}

Due to the nature of our proposed pruning methods, they can be applied using any pruning criterion. In this section, we compare the coding performance of our proposed pruning methods on three pruning criteria: L2-Norm of the weights \cite{li2016pruning}, Rank of feature maps (HRANK) \cite{lin2020hrank} and Channel Independence (CHIP) \cite{sui2021chip}. Since HRANK and CHIP are feature-guided pruning criteria, we use $10$ images from our training set to calculate these metrics. We also create a calibration dataset consisting of $250$ images from the COCO training set and use it for the NAS method. Once the layer-wise pruning ratio has been obtained using our NAS procedure, we perform a finetuning of the pruned model for $60$ epochs with a learning rate of $10^{-4}$.

We observe from Table.~\ref{tab:tab1} that simultaneously pruning filters and filter channels as opposed to only pruning filters can achieve a BD-Rate reduction of $11.63\% (42.19-30.56)\%$, $26.26\% (34.09-7.83)\%$ and $8.9\% (7.17+1.73)\%$ using the L2-Norm, HRANK and CHIP criteria respectively under similar model size constraints. We can also see that the improvement in BD-Rate is even more ($31.86\% (42.19-10.33)\%$, $33.72\% (34.09-0.37)\%$ and $9.17\% (7.17+2.0)\%$ using the L2-Norm, HRANK, and CHIP criteria respectively) when we incorporate our NAS procedure to find the layer-wise pruning ratio instead of assigning a fixed pruning ratio to each layer. Therefore, our proposed pruning methods can improve rate-distortion performance for a given model size constraint.

\begin{table}[t]
\begin{center}
\caption{Trade-off between BD-Rate and model compression achieved by our proposed pruning methods using the CHIP criterion.} \label{tab:tab2}
\vspace{-0.3cm}
\scalebox{0.91}{
\begin{tabular}{c|c|c|c}
\hline
\multirow{2}{*}{\begin{tabular}[c]{@{}c@{}} \\ LIC \\ Model \end{tabular}} & \multirow{2}{*}{\begin{tabular}[c]{@{}c@{}} \\  $S_{\text{target}}$\\(\%)\end{tabular}} & \multicolumn{2}{c}{BD-Rate(\%)} \\
\hhline{~~--}
& & \begin{tabular}[c]{@{}c@{}}Fixed \\Pruning Ratio\end{tabular} & \begin{tabular}[c]{@{}c@{}}Layer-wise \\Pruning ratio\\ (NAS) \end{tabular} \\
\hline
\multirow{3}{*}{\begin{tabular}[c]{@{}c@{}}Mean Scale\\ Hyperprior \cite{minnen2018joint}\end{tabular}} & 30 & -1.73      & -2.00 \\
& 45 & +5.38      & +4.48 \\ 
& 60 & +11.86     & +9.87 \\ \hline
\end{tabular}}
\vspace{-0.4cm}
\end{center}
\end{table}

\begin{table}[t]
\begin{center}
\caption{Coding performance of our proposed joint pruning and quantization method using the Mean Scale Hyperprior \cite{minnen2018joint} model.} \label{tab:tab3}
\vspace{-0.3cm}
\scalebox{0.95}{
\begin{tabular}{c|c|c|c}
\hline
\begin{tabular}[c]{@{}c@{}} Compression \\ Type \end{tabular} & 
\begin{tabular}[c]{@{}c@{}} BD-Rate \\ (\%) \end{tabular} & 
\begin{tabular}[c]{@{}c@{}} Model Size \\ (MB) \end{tabular} & 
\begin{tabular}[c]{@{}c@{}} Compression \\ Ratio \end{tabular} \\
\hline
Pruned (80\%) & +29.35 & 5.97 & $4.6\times$ \\
\hline
\begin{tabular}[c]{@{}c@{}} Pruned (20\%) + \\ Quantized \end{tabular} & +7.39 & 5.57 & $4.9\times$ \\
\hline
\begin{tabular}[c]{@{}c@{}} Pruned (35\%) + \\ Quantized \end{tabular} & +8.12 & 4.23 & $6.6\times$ \\ 
\hline
\end{tabular}}
\vspace{-0.6cm}
\end{center}
\end{table}


\vspace{-0.2cm}
\subsection{Trade-off between BD-Rate and Model Size}
\label{sec:exp_3}
\vspace{-0.2cm}

Our model can achieve a trade-off between BD-Rate and parameter reduction by varying the hyperparameter $S_{\text{target}}$. We demonstrate this in Table.~\ref{tab:tab2}. With $S_{\text{target}}$ set to $30\%$, our pruned model using the NAS procedure can achieve a BD-Rate of $-2.0\%$ yielding better rate-distortion performance compared to the uncompressed baseline model. As $S_{\text{target}}$ is increased from $30\%$ to $45\%$ and $60\%$, the BD-Rate performance drops from $-2.0\%$ to $+4.48\%$ and $+9.87\%$ while achieving greater model size reductions. Therefore, our proposed model can trade-off between model size and rate-distortion performance. Our proposed model without the NAS procedure using fixed pruning ratios also produces similar trends. As $S_{\text{target}}$ is increased from $30\%$ to $45\%$ and $60\%$, we get increased model compression while BD-Rate drops from $-1.73\%$ to $+5.38\%$ and $+11.86\%$

From Table.~\ref{tab:tab2}, we can further see that incorporating our NAS method to find the layer-wise pruning ratio (last column) consistently outperforms assigning fixed pruning ratios (second last column) to each layer of the network for models pruned to different pruning ratios. Furthermore, as the model becomes more sparse, the gap between the performance gain due to our NAS method becomes more prominent. For this experiment, we simultaneously prune filters and filter channels as described in Sec.~\ref{sec:method_1} and use the CHIP pruning criterion. We also set $\delta$ defined in Sec.~\ref{sec:method_2} to $0.01$. 

\vspace{-0.3cm}
\subsection{Coding Performance of Pruning + Quantization}
\label{sec:exp_4}
\vspace{-0.2cm}

As can be seen from the last row of Table.~\ref{tab:tab2} and first row of Table.~\ref{tab:tab3}, pruning too many parameters from the model significantly degrades its rate-distortion performance (at $60\%$ and $80\%$ sparsity BD-Rate drops to $9.87\%$ and $29.35\%$ respectively). This may be because a highly pruned model does not have sufficient parameters to capture all the information content of the image. To overcome this problem, we prune our model to a moderate sparsity and then achieve additional model compression by quantizing our pruned model. We use the quantization method described in Sec.~\ref{sec:method_3} to quantize our pruned network to 8 bits. In our experiments, we first prune the model using our proposed method. We then finetune the model for $20$ epochs before quantizing it and finetuning again for $40$ epochs.

\begin{figure}[t]
 \centerline{\epsfig{figure=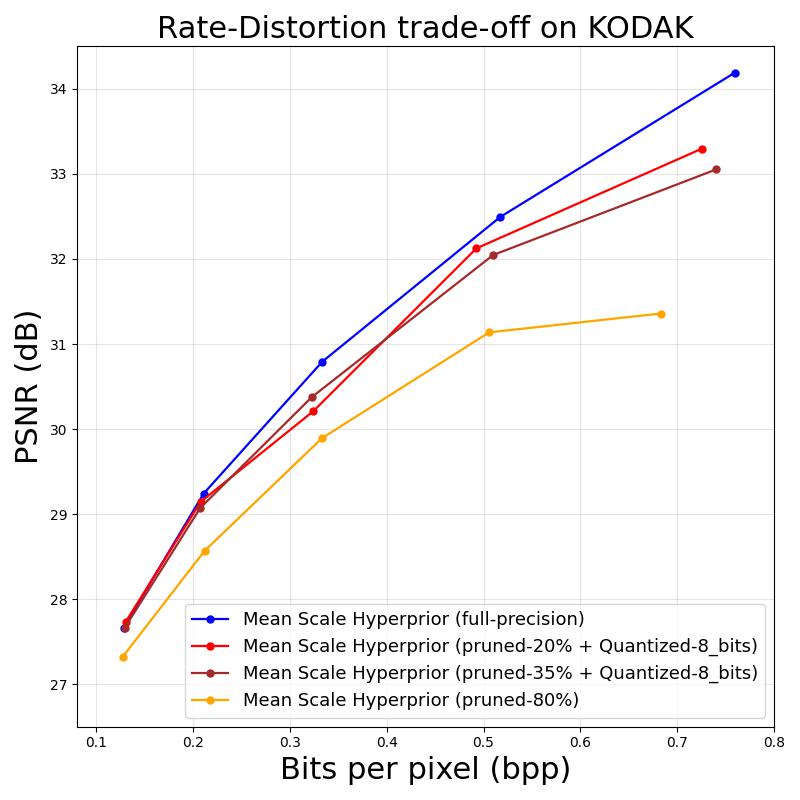, width=0.85\linewidth}}
 \vspace{-0.3cm}
\caption{Comparison of the rate-distortion curve of our baseline LIC model with those of our pruned and quantized models using the KODAK dataset.}
\vspace{-0.4cm}
\label{fig:rd_curve}
\end{figure}

Table.~\ref{tab:tab3} demonstrates the effectiveness of performing quantization in conjunction with pruning. If we prune $20\%$ of model parameters and then quantize to 8 bits, we can achieve $4.9 \times$ model compression while maintaining a BD-Rate of $+7.39\%$. On the contrary, pruning $80\%$ of the network without quantization achieves similar model size reduction ($4.6 \times$), but BD-Rate performance drops to $+29.35\%$. In fact, even if we prune $35\%$ of the model and then quantize to 8 bits, we can obtain a better BD-Rate performance ($+8.12\%$) than pruning $80\%$ of the model while achieving a higher model size reduction ($6.6 \times$). The rate-distortion curves for this experiment are shown in Fig.~\ref{fig:rd_curve}.

\vspace{-0.2cm}





\section{Conclusion}
\label{sec:conclusion}
\vspace{-0.3cm}

In this work, we propose a model compression scheme targeted for LIC models based on structured pruning and quantization. We perform experiments to show that simultaneously pruning filters and filter channels can produce better rate-distortion performance than only pruning filters under the same model size constraints. We also formulate a Neural Architecture Search (NAS) procedure that can determine the layer-wise pruning ratio for the pruned LIC model while conforming to a target model size requirement. Our experiments show that the NAS procedure can enhance rate-distortion performance compared to assigning a fixed pruning ratio for each layer. We finally find that pruning too many parameters of the LIC model can significantly degrade performance as the network no longer has sufficient parameters to accurately characterize the image. However, we mitigate this problem by pruning our LIC to a moderate value, and then quantizing it to an 8-bit fixed-precision network to achieve the necessary model compression. We show that this joint pruning and quantization scheme yields significantly better results than only pruning the network.

\bibliographystyle{IEEEbib}
{\footnotesize \bibliography{icip2024main}}

\end{document}